Evidence of Water-related Discrete Trap State Formation in Pentacene Single Crystal Field-Effect Transistors

C. Goldmann, D. J. Gundlach, and B. Batlogg

Laboratory for Solid State Physics

ETH Zurich, 8093 Zurich, Switzerland

Phone: + 41 1 633 2361, Fax: + 41 1 633 1072, Email: goldmann@solid.phys.ethz.ch

*Abstract* – We report on the generation of a discrete trap state during negative gate bias stress in pentacene single crystal "flip-crystal" field-effect transistors with a $SiO_2$ gate dielectric. Trap densities of up to $2 \cdot 10^{12}$ cm$^{-2}$ were created in the experiments. Trap formation and trap relaxation are distinctly different above and below ~280 K. In devices in which a self-assembled monolayer on top of the $SiO_2$ provides a hydrophobic insulator surface we do not observe trap formation. These results indicate the microscopic cause of the trap state to be water adsorbed on the $SiO_2$ surface.

Organic semiconductors are of technological interest for large-area or low-cost electronics. Whereas organic thin-film transistors (OTFTs) are used in applications, research on single crystals elucidates the intrinsic transport properties and their extrinsic limitations. Even though organic semiconductors have been studied for decades, relatively little is known about traps in these materials. Bias stress measurements can provide insight on charge trapping.[1-5] For example, the formation of a hydrogen-related metastable hole trap in pentacene under bias stress has been reported.[6] An important aspect of charge trapping in organic semiconductors is the influence of ambient conditions.[7,8] Of equal importance in field-effect transistors (FETs) is the effect of the gate dielectric which may contain or induce additional traps.[9]

We have discovered and studied the formation of a discrete hole trap after negative bias stress in pentacene single crystal FETs with a $SiO_2$ gate dielectric. The density of traps formed and their relaxation is distinctly different above and below ~280 K. A comparison to FETs with a hydrophobic dielectric surface indicates the microscopic cause of the trap state to be related to molecular layers of water adsorbed on the dielectric surface.

Pentacene "flip-crystal" field-effect devices were fabricated on oxidized Si wafers according to the technique reported on previously.[5,10] A schematic top view of the device layout is shown in the inset of Fig. 1. The oxidized silicon substrates were cleaned in solvents, sulphuric peroxide, and thoroughly rinsed with DI water. This procedure renders the $SiO_2$ surface hydrophilic, with a few monolayers of adsorbed water ([11] and references therein). For comparison, some substrates were subsequently treated with octadecyltrichlorosilane (OTS), which is known to form self-assembled monolayers on the dielectric surface rendering it hydrophobic.[12]

Guarded measurements were carried out in darkness and in helium atmosphere using a HP 4155A Semiconductor Parameter Analyzer with an expander unit (HP 41501A).

The normalized I-V characteristics of different pentacene FETs on untreated $SiO_2$ after negative bias stress are shown in Fig. 1. The hole mobility of the FETs in the linear regime (2-terminal measurement) was in the range of 0.1-0.5 $cm^2$/V·s, and as high as 1.5 $cm^2$/V·s in one sample when extracted using the four-terminal technique. The step-like features in the I-V curves of the stressed devices result from the formation and subsequent filling of a discrete trap state. Out of a dozen samples, five showed a pronounced step-like feature like that plotted in Fig. 1 (filled symbols), and another five showed a decreased subthreshold slope (open symbols). In two samples we did not find indications of trap formation. The formation of the trap state we report was not influenced by illumination of the sample with microscope light during stress.

Trap formation was studied at different temperatures. Fig. 2(a) shows the evolution with time of the trap state at 297 K. Initially the trap density increases quickly, and saturates after a few hrs. To quantify the induced trap density we define two characteristic voltages $V_{o1}$ and $V_{o2}$ as those values of $V_{GS}$ where $I_D$ exceeds $I_{o1} = 5·10^{-7}$ A and $I_{o2} = 5·10^{-10}$ A, respectively (Fig. 2(a)). $I_{o1}$ and $I_{o2}$ correspond to points of the I-V curve below and above the trap-filling "step", where we expect the quasi-Fermi level (qFL) to lie below and above the discrete trap level. The evolution of the discrete trap density is determined from the gate capacitance ($1·10^{-8}$ F/$cm^2$) and the difference between $V_{o2}$ - $V_{o1}$ after a given stress time and $V_{o2}$ - $V_{o1}$ before stress. This is shown in the inset to Fig. 2(b) for T = 268 K and 297 K. Interestingly, the characteristic times for trap formation ($\cong$100 min) are not significantly temperature-dependent. In particular, these characteristic times are considerably higher than the time required to measure a transfer characteristic (typically ~15 s). After ~5 hrs of stress the trap density has nearly reached its saturation value at each temperature. For a quantitative analysis we could either take the value at 5 hrs, or, as we did, the saturation value derived from an exponential fit to the data.[13] The difference is minor and does not affect the main conclusion. The saturation density for the different temperatures is plotted in Fig. 2(b). Trap formation at 297 K was measured twice (before and after the measurements at the other temperatures), and shows good reproducibility after tens of hrs of stress. At 310 K, 297 K, and 288 K the trap density saturates at $\sim 1.8 \cdot 10^{12}$ cm$^{-2}$, while it only reaches $\sim 0.4 \cdot 10^{12}$ cm$^{-2}$ for 279 K and 268 K. We find this remarkable given the relatively small change in temperature from 288 K to 279 K.

To monitor trap relaxation the device reported in Fig. 3 was stressed at 297 K for one hour ($V_{GS} = -60$ V) and then cooled to lower temperature with $V_{GS}$ still applied. The time for cooling to 245 K was 60 min and that to 265, 270, 280, and 284 K was 40 min. As shown in Fig. 3(a) for T = 265 K it was possible to preserve the large trap density created at 297 K when cooling to lower temperatures. We characterize the relaxation of the discrete trap state using the current levels shown in Fig. 3(a). The time evolution of the voltage difference $V_{o2} - V_{o1}$ as a measure of the trap density is plotted in Fig. 3(b) for selected temperatures. The decay dynamics show a pronounced temperature-dependence over the relatively narrow temperature range between 245 K and 297 K. At T ≤ 270 K the decay can be fitted by the sum of two exponentials indicative of two distinct time scales. The "slow" process rapidly speeds up with increasing temperature, and for T ≥ 280 K the trap completely relaxes in less than 5 min described by a single characteristic time which is close to the short timescale for T ≤ 270 K. Interestingly, this temperature range 270 K ≤ T ≤ 280 K is similar to the one where we observe a distinct change in trap density for trap formation. The fast relaxation of the trap state above 280 K indicates that it is of a different origin than the trap observed by Lang et al. for free standing pentacene crystals, who report relaxation times of ~1 hr at room temperature in darkness.[6]

To identify the microcopic cause of the discrete trap we have also carried out bias stress studies on devices prepared on $SiO_2$ that was rendered hydrophobic by an OTS treatment (Fig. 4). No indication of trap generation under negative bias stress was observed. In addition, one of the samples fabricated on untreated $SiO_2$ originally showed pronounced trap formation, but the step-like feature could no longer be created after 2 weeks of storage in helium. However, trap generation could be reactivated by placing the sample in room air for ~18 hrs. We therefore conclude that the trap state formation is related to molecular layers of water adsorbed on the $SiO_2$ dielectric surface.

Jurchescu et al.[8] reported that humid air diffuses into pentacene crystals and that the water molecules cause charge trapping. Here, we observe trap formation only after negative bias stress. While the exact mechanism of trap formation remains to be identified, we note that the timescales observed here are typical for diffusion of small ions or molecules in organic single crystals[6,8] and suggest a diffusion-limited chemical reaction. The marked difference of trap formation above and below 280 K may be associated with a phase transition in the adsorbed monolayers of water. If due to "freezing", the difference between ~280 K and the bulk freezing at 273 K may not be unusual. Such phenomena in mesoscopic systems are well studied.[14,15] We note here that phase transitions of water may play a role in other organic devices as well, cf. [3].

With respect to the position in energy of the qFL during trap formation we found for the device shown in Fig. 3 that the trap state only appeared after stress at -60 V but not after stressing at -40 V for even longer periods of time. This indicates that for trap formation the qFL has to be moved past a certain level in the gap. The voltage necessary for this shift obviously depends on the over-all trap density in a given crystal. This is discussed for a specific case in [6]. Furthermore, even if the trap state is created it can be masked by a high "background trap density" as one might expect in OTFTs.

We have found and studied the formation of a discrete trap level in pentacene FETs associated with adsorbed water at the gate dielectric surface. Trap generation occurs when the quasi-Fermi level is moved sufficiently close to the valence band and apparently requires a few molecular layers of water at the interface. It is not observed when the gate dielectric is rendered hydrophobic by an OTS treatment. Both trap formation and trap relaxation are distinctly different above and below ~280 K. This study emphasizes the crucial role that adsorbed water may play in organic field-effect devices in general as they are commonly fabricated.

We thank K. P. Pernstich for assistance with the measurements, and S. Haas for help with crystal growth.

**References**

1. R. A. Street, A. Salleo, and M. L. Chabinyc, Phys. Rev. B **68**, 085316 (2003). A. Salleo and R. A. Street, Phys. Rev. B **70**, 235324 (2004). A. Salleo, F. Endicott, and R. A. Street, Appl. Phys. Lett. **86**, 263505 (2005).

2. M. Matters, D. M. de Leeuw, P. T. Herwig, and A. R. Brown, Synth. Met. **102**, 998 (1999).

3. H. L. Gomes, P. Stallinga, F. Dinelli, M. Murgia, F. Biscarini, D. M. de Leeuw, T. Muck, J. Geurts, L. W. Molenkamp, V. Wagner, Appl. Phys. Lett. **84**, 3184 (2004).

4. D. Knipp, R. A. Street, A. Völkel, and J. Ho, J. Appl. Phys. **93**, 347 (2003).

5. C. Goldmann, C. Krellner, K. P. Pernstich, S. Haas, D. J. Gundlach, and B. Batlogg, submitted.

6. D. V. Lang, X. Chi, T. Siegrist, A. M. Sergent, and A. P. Ramirez, Phys. Rev. Lett. **93**, 076601 (2004).

7. C. R. Kagan, A. Afzali, and T. O. Graham, Appl. Phys. Lett. **86**, 193505 (2005).

8. O. D. Jurchescu, J. Baas, and T. T. M. Palstra, Appl. Phys. Lett. **87**, 052102 (2005).

9. L.-L. Chua, J. Zaumseil, J.-F. Chang, E. C.-W. Ou, P. K.-H. Ho, H. Sirringhaus, and R. H. Friend, Nature **434**, 194 (2005).

10. J. Takeya, C. Goldmann, S. Haas, K. P. Pernstich, B. Ketterer, and B. Batlogg, J. Appl. Phys. **94**, 5800 (2003). C. Goldmann, S. Haas, C. Krellner, K. P. Pernstich, D. J. Gundlach, and B. Batlogg, J. Appl. Phys. **96**, 2080 (2004).

11. A. C. Mayer, R. Ruiz, R. L. Headrick, A. Kazimirov, and G. G. Malliaras, Org. Electron. **5**, 257 (2004).

12. Y. Y. Lin, D. J. Gundlach, S. F. Nelson, and T. N. Jackson, IEEE Electr. Dev. Lett. **18**, 606 (1997).

13. A single exponential was usually sufficient.

14. E.-M. Choi, Y.-H. Yoon, S. Lee, and H. Kang, Phys. Rev. Lett. **95**, 085701 (2005).

15. Y. Maniwa, H. Kataura, M. Abe, S. Suzuki, Y. Achiba, H. Kira, and K. Matsuda, J. Phys. Soc. Jpn. **71**, 2863 (2002).

**Figure Captions**

Fig. 1  Transfer characteristics ($V_{DS}$ = -10 V) for different pentacene FETs after negative bias stress ($V_{GS}$ = -60 V, 60 min). For comparison the unstressed characteristic of one of the devices is also shown.

Fig. 2  (a) Transfer characteristics ($V_{DS}$ = -10 V) showing trap formation at 297 K for stress times of 0, 15, 35, 60, ~107, 160, 220, and 280 min ($V_{GS}$ = -60 V, sample B).

(b) Saturation value of the discrete trap density at different temperatures. Inset: Time evolution of $V_{o2}$ - $V_{o1}$ (corrected for the "unstressed" subthreshold slope) at 297 K and 268 K.

Fig. 3  (a) Trap relaxation dynamics (0 min, ~1 min, ~6 min, ~ 11 min, and ~ 32 min, $V_{DS}$ = -10 V, sample A) after the device was stressed at $V_{GS}$ = -60 V for 1 h at 297 K, and then cooled to 265 K with the stress applied.

(b) Corresponding $V_{o2}$ - $V_{o1}$ vs. time for relaxation at 245 K (filled diamonds), 265 K (open circles), 270 K (filled triangles), and 284 K (open squares).

Fig. 4  Comparison between devices with a standard $SiO_2$ and with an OTS-treated hydrophobic dielectric surface ($V_{GS}$ = -60 V for 100 min at room temperature).

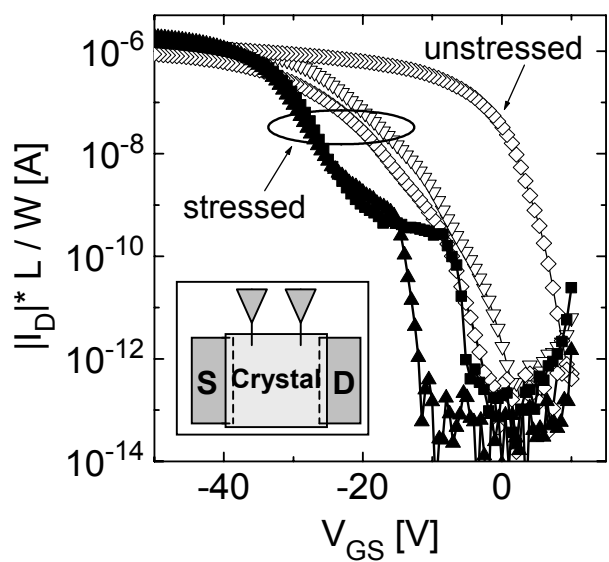

*Fig. 1 of 4, C. Goldmann et al.*

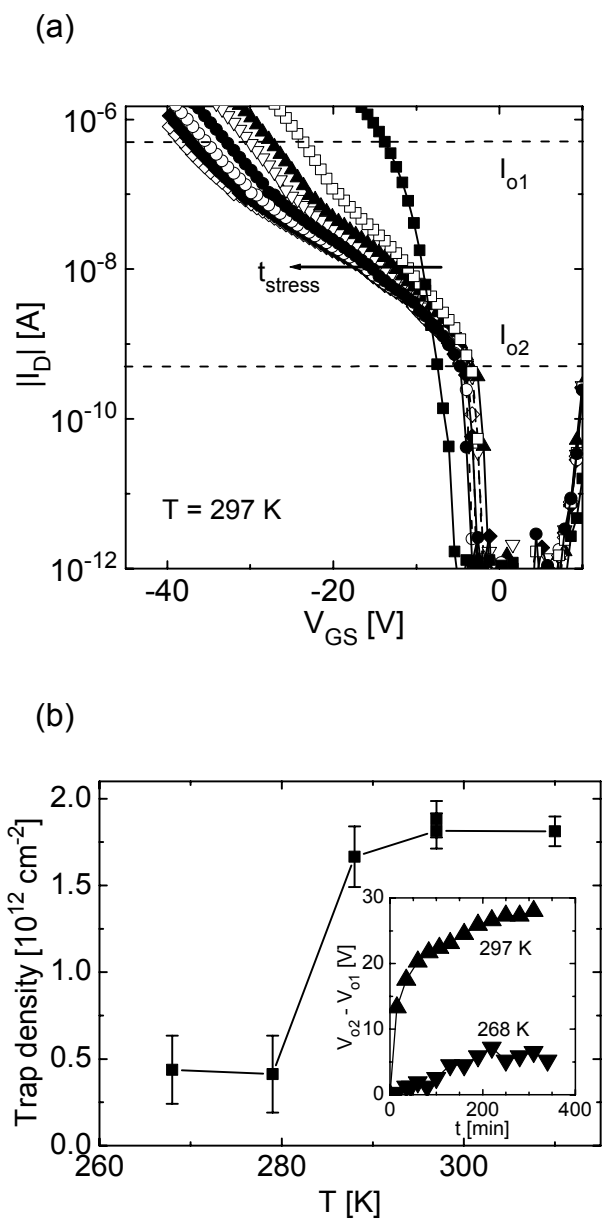

*Fig. 2 of 4, C. Goldmann et al.*

(a)

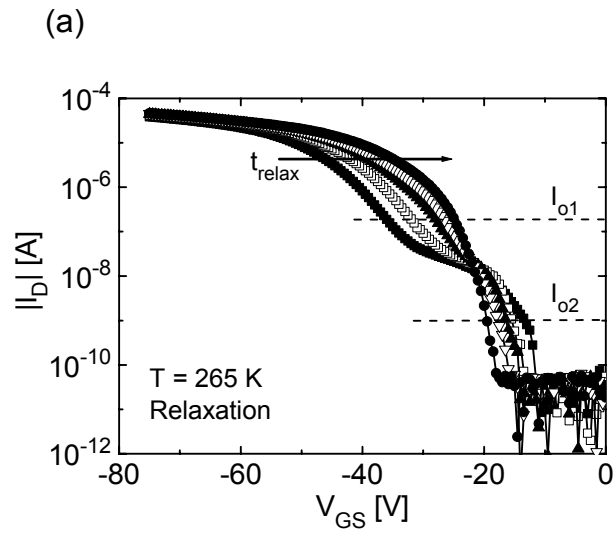

(b)

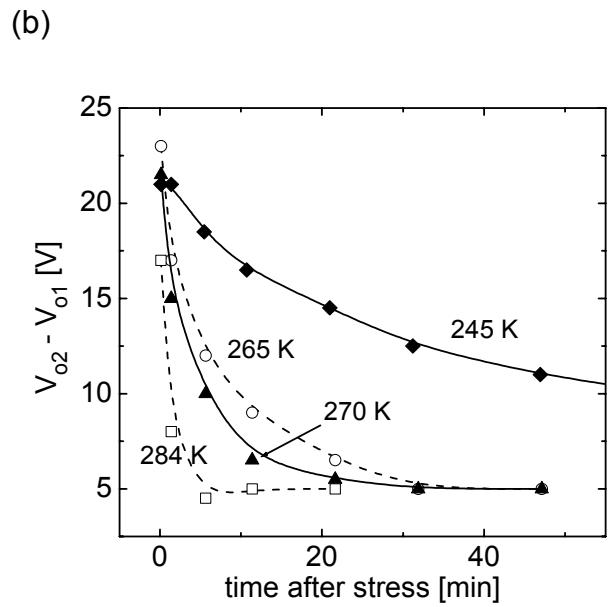

*Fig. 3 of 4, C. Goldmann et al.*

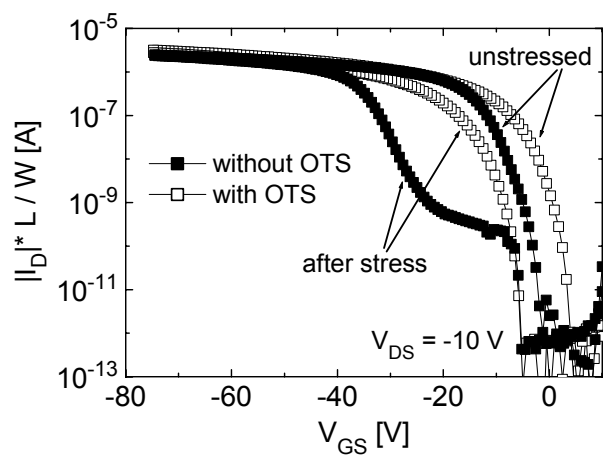

*Fig. 4 of 4, C. Goldmann et al.*